\documentclass[onefignum,onetabnum]{siamart190516} 

\usepackage{lipsum}
\usepackage{amsfonts}
\usepackage{graphicx}
\usepackage{epstopdf}
\usepackage{algorithmic}
\ifpdf
  \DeclareGraphicsExtensions{.eps,.pdf,.png,.jpg}
\else
  \DeclareGraphicsExtensions{.eps}
\fi

\newsiamremark{remark}{Remark}
\newsiamremark{hypothesis}{Hypothesis}
\crefname{hypothesis}{Hypothesis}{Hypotheses}
\newsiamthm{claim}{Claim}

\headers{Speeding up Python-based Lagrangian Particle Simulations}{C. Kehl, E. van Sebille and A. Gibson}

\title{Speeding up Python-based Lagrangian Fluid-Flow Particle Simulations via Dynamic Collection Data Structures\thanks{Submitted to the SIAM Journal on Scientific Computing editors 2021-01-25.
\funding{This work was funded by the European Research Council (ERC) under TOPIOS project (grant no.~715386). Simulations were carried out on the SURF's Dutch National e-Infrastructure (project no. 16371 and 2019.034).}}} 

\author{Christian Kehl
\and Erik van Sebille\thanks{Institute for Marine and Atmospheric Research, Utrecht University, the Netherlands 
  (\email{c.kehl@uu.nl}, \email{e.vansebille@uu.nl}).}
\and Angus Gibson\thanks{The Australian National University (\email{angus.gibson@anu.edu.au}).}
}

\usepackage{amsopn}

\ifpdf
\hypersetup{
  pdftitle={Speeding up Python-based Lagrangian Fluid-Flow Particle Simulations via Dynamic Collection Data Structures}
  pdfauthor={C. Kehl, E. van Sebille and A. Gibson}
}
\fi

\definecolor{newcolor}{rgb}{.8,.349,.1}

\RequirePackage[UKenglish]{babel}
\graphicspath{{./}}
\DeclareGraphicsExtensions{.jpg,.jpeg,.png,.eps}
\RequirePackage[center,tight,footnotesize]{subfigure}
\RequirePackage[acronyms,shortcuts]{glossaries}
\RequirePackage[normalem]{ulem}
\makeglossaries

\newacronym{AoS}{AoS}{array-of-structures}
\newacronym{SoA}{SoA}{structure-of-arrays}
\newacronym{LoA}{LoA}{list-of-arrays}
\newacronym{LoAoS}{LoA(oS)}{list-of-array-of-structures}
\newacronym{CFD}{CFD}{computational fluid dynamics}

\begin{document}

\maketitle

\begin{abstract}
  Array-like collection data structures are widely established in Python's scientific computing-ecosystem for high-performance computations. The structure maps well to regular, gridded lattice structures that are common to computational problems in physics and geosciences. High performance is, however, only guaranteed for static computations with a fixed computational domain. We show that for dynamic computations within an actively changing computational domain, the array-like collections provided by NumPy and its derivatives are a bottleneck for large computations. In response, we describe the integration of naturally-dynamic collection data structures (e.g. double-linked lists) into NumPy simulations and \textit{ctypes}-based C-bindings. Our benchmarks verify and quantify the performance increase attributed to the change of the collection data structure. Our application scenario, a Lagrangian (oceanic) fluid-flow particle simulation within the \textit{Parcels} framework, demonstrates the speed-up yield in a realistic setting and demonstrates the novel capabilities that are facilitated by optimised collection data structures.
\end{abstract}



\section{Introduction}
During the past decade, the rise of Python as a versatile programming platform for scientific research, particularly in physics and geoscience, has led to the development of various domain-specific numerical simulation- and modelling platforms for subsurface modelling (e.g. GemPy~\cite{Varga2019}), plate tectonics (e.g. GPlate~\cite{Mueller2018}), Eulerian oceanic flow simulations (e.g VEROS~\cite{Haefner2018} and others~\cite{Lemenkova2019}) and Lagrangian particle tracing within ocean currents (e.g. Parcels~\cite{Delandmeter2019a}). Simulation and modelling with Python is advantageous for its coding simplicity for novice developers and domain experts, its versatility due to its typeless, interpreter-based processing and its computational efficiency via its links to establish high-performance linear algebra libraries (e.g. BLAS~\cite{Blackford2002} and LAPACK~\cite{Anderson1999a}) through packages such as NumPy~\cite{VanderWalt2011,Harris2020}, SciPy~\cite{Virtanen2020} and Scikits~\cite{VanderWalt2014,Pedregosa2011}. Additionally, Python's embedding in modern DevOps tools (e.g. Git) and its wide application in open-source and open-science projects are additional reasons for its high acceptance in the scientific community.

The development of Parcels~\cite{Lange2017,Delandmeter2019a}, which is our application platform in this paper, relies on the high-performance computation within Python. In this regard, it already employs \textit{ctypes}~\cite{Oliphant2007} as just-in-time (JIT) C-interface to execute highly-adaptive kernels for the advection and tracing of particles on ocean general circulation model (OGCM) fields. Those kernels are provided as elementary Python functions by the developers as well as the user community, and are compiled at runtime into C-code for rapid field interpolation and advection evaluation. The available high-performance packages, such as NumPy, handle dense-matrix data organisation with high efficiency. On the other hand, problems that cannot be translated into a dense-matrix algebra problem, or data that are dynamically altered during the simulation, pose a challenge to efficient calculations, because the re-organisation of the data container (i.e. \textit{array-like} collection) can in some cases consume more runtime as memory overhead than is required for the particle advection itself. This issue also arises in other physics simulations that concern large sets of dynamically-changing physics entities (e.g. the common \textit{nBody} simulation where bodies are 'on-the-fly' added and deleted).


After careful benchmarking (\cref{sec:statusquo}), we identified this data collection re-organisation as the major barrier to increased performance for particle simulations in fluid flows. In this paper, we propose the introduction of alternative data structures for the organisation of dynamic data entities in Python. Furthermore, those data structures are expressed in a JIT-compliant format for increased performance. The major contribution of this article is a detailed performance evaluation and comparison of various data structures and data organisation patterns. 

The starting point of our investigation is a common array-of-structure arrangement of the particles within a dense array. The performance benefit of a structure-of-array arrangement, which allows for higher cache coherence and better data locality in vectorized computations, is evident for static data arrays. On the other hand, a list-of-array arrangement with  sublists of defined maximum size, limiting the memory reordering overhead to a fixed upper bound, only provides marginal improvements despite the theoretical upper bound on computational complexity (see \cref{sec:method:loa}). In scenarios with dynamic data arrays, where particle entities are inserted-into and removed-from the particle collection arbitrarily during the simulation, the runtime expense for data reordering, memory copies and poor data locality becomes evident. In that context, double-linked list containers show a distinct performance advantage, especially for large particle collections with more than $2^{15}$ entities. 

\section{Related Work}
\label{sec:relwork}

In this section, we first review the current state of efficient (collection) data structures in Python's scientific computing environment. As this is a comprehensive topic with abundant literature, we concentrate mainly on non-array-like collections and their access potential in Python -- C wrappers.

Afterwards, we present the background knowledge on our physics application of Lagrangian oceanic flow simulations via particles. That section mainly serves as an introduction to our workflow and provides background information to the performance- and runtime bottleneck concerns.

\subsection{Efficient data structures in scientific computing}
\label{sec:relwork:efficient_ds}

Native Python is known for its hampered processing performance. In order to circumvent the compute bottleneck, NumPy~\cite{VanderWalt2011} was introduced as fast C-wrapper to BLAS and LAPACK, which arranges its data efficiently as blockwise \textit{matrices} or \textit{dense arrays} (i.e. array collections). Those \textit{array-like} collections, such as dense arrays and matrices, are common high-performance data structures, especially in Python, for modern scientific computing. The data structure's main objective, which is random access of items, is rarely required in modern computing. Frameworks such as SciPy~\cite{Virtanen2020}, scikit-image~\cite{VanderWalt2014}, scikit-learn~\cite{Pedregosa2011} and other machine-learning frameworks \cite{Bergstra2010a,Abadi2016a,Chollet2018}) build on top of NumPy's dense-array and matrix data structures. A common, fundamental issue with \textit{array-like} collections is that random insertions and deletions of items in the collection requires array copies and merges (see \cite{Sedgewick1998}). This slows down the operation and requires an additional data buffer of the size of the collection, which can be prohibitively expensive in modern big data-computing.

The Python-internal \textit{bisect} module facilitates a more \textit{list-like} collection behaviour via binary searches of objects in \textit{array-like} collections. That said, the binary search only simplifies object search and object access, but it provides no functionality to speed-up native, internal item insertion or deletion. This functionality is provided by the \textit{collections.deque} module, which is a linked nodelist. In terms of item insertions or deletions performance, it exhibits the expected constant runtime complexity (i.e. logarithmic complexity, when including item localisation). That said, working with the \textit{collections.deque} module is difficult because the C-interface and data structure access (e.g. via \textit{ctypes}) is poorly documented. Moreover, the connection cannot reliably be established because the C-interface is a double-linked list whereas the Python interface is a uni-linked list.

The \textit{sortedcollections} is another Python module that promises a C-comparable performance behaviour for dynamic collections, such as the \textit{SortedList} for 'std::list' equivalence. The module internally wraps Python \textit{lists} (i.e. array-like collections) in an internally-managed item indexing scheme. The indexing scheme provides constant-complexity insertions and deletions, while the use of the \textit{bisect} module internally facilitates fast object access. All together, the module provides all performance-relevant features and advantages of \textit{list}-like collections. The drawback of the module is the lack of Python -- C interfaces for fast execution of simulations and analyses.

The gap we address in this paper is the development and benchmark of dynamic collection data structures in Python with an integrated C-interface for rapid computations on dynamically-organised data.

\subsection{Lagrangian Particle Tracing in Oceanic Flows with Parcels}
\label{sec:relwork:parcels}

Our physics-motivated prime application scenario is the tracking of (virtual) particles in oceanic flows. These virtual particles can for example represent plastic~\cite{Lebreton2012}, plankton~\cite{Nooteboom2019}, nutrients~\cite{Cetina-Heredia2018} or fish~\cite{ScuttPhillips2018}. The tracing is done by Lagrangian advection, where particles follow a given background (Eulerian) advective flow field \cite{Delandmeter2019a} and whose trajectories are numerically integrated over time as given in eq. \cref{eq:advection}, implemented in e.g., an Euler-forward- or Runge-Kutta scheme~\cite{Nordam2020}.

\begin{equation}\label{eq:advection}
X(t+\Delta t) = X(t) + \int_{t}^{t+\Delta t} v(x(\tau), \tau) d \tau + dX_b,
\end{equation}

with $X(t) = x$ representing an individual particle's position at time $t$, advected in $\Delta t$ timesteps and discretized over $\tau$, $v(x(t))$ is the (gridded) fluid flow from a numerical simulation, and $dX_b$ is an optional position change because of particle `behaviour' (swimming, sinking, etc). The partial integral is required to match up the temporal resolutions of the underlying flow field and the particle simulation.

Considering the computational complexity in this particle advection scheme, there is a distinction between the \textit{theoretical} and \textit{practical} computational workload. Theoretically, the operations that majorly contribute to the compute load are (i) the velocity interpolation $\int_{t}^{t+\Delta t} v(x(\tau), \tau) d \tau$ on the gridded Eulerian flow field, (ii) the position integration $X(t+\Delta t)$ with a fixed- or adaptive timestepping and (iii) the sum of partial integration results (i.e. \textit{fused multiply-add (fma)} operations). Next to those plain calculations for the particle trajectory, the Lagrangian simulation's main purpose is the tracking of particle properties (be they of physical, biochemical or behavioural nature) over time. Those functions are executed in Parcels by the help of user-defined \textit{kernels}. Depending on the complexity of the particle model, those operations represent the majority of calculations in a \textit{practical} setting. On the other hand, as those kernels are case-specific and hardly generalizable, their compute workload is not considered in this paper's experiments.

In terms of performance and benchmarking, the \textit{runtime} of the whole simulations, as well as each integration iteration individually, is the sum of the compute time (as analysed above) and the input-/output (I/O) time. The latter is further split into memory-I/O, file-I/O and plotting time (which is the special case for device output happening not exclusively to files nor in-memory). 

In realistic, practical Lagrangian fluid flow simulations, the compute time is by far exceeded by the file-I/O, as the (oceanic) flow fields need to be dynamically loaded from large file databases. Hence, from a practical perspective, Lagrangian fluid flow simulations are typically strictly \textit{I/O-bound} tasks. Parcels \cite{Delandmeter2019a} already implements multiple I/O-enhancing techniques (e.g. deferred loading, dynamic and selective loading of exclusively occupied cells), and thus we are hereby more concerned about improving the particle integration itself for large particle sets (i.e. \textit{particle swarms}). Therefore, all experiments in this paper utilize synthetic, pre-loaded perlin-noise flow fields \cite{Perlin1985} that remain resident in memory over the whole simulation, neglecting file-I/O time of the flow fields.

An in-depth performance analysis is required for the case of large particle sets, because effects from hardware caching, memory layout and particle set data organisation have a (potentially) measurable impact on the overall runtime. From the perspective of Lagrangian fluid flow simulations as a \textit{particle system}, a particle set can be \textit{statically} allocated and remains fixed in its data entities over the the whole simulation. Alternatively, in order to circumvent limited-precision floating-point errors and to comply with case-specific simulation requirements, particles can be \textit{dynamically} re-emitted and removed from the particle set. Both cases (potentially) behave differently in terms of performance and runtime due to data access- and alteration patterns for different collection data structures that can be used to store a particle set.

The potential performance difference between static and dynamic particle sets, its quantification, its rationale, its impact on different oceanic flow simulation cases, and its exploitation for speed improvement are the main experimental subjects of this article.

\section{Status-Quo Benchmarks}
\label{sec:statusquo}

\begin{figure}[htbp]
	\begin{center}
	 	\begin{minipage}{\columnwidth}
        \centering
			\subfigure[dynamic particle insertion]
			{\includegraphics[keepaspectratio, width=0.485\columnwidth]{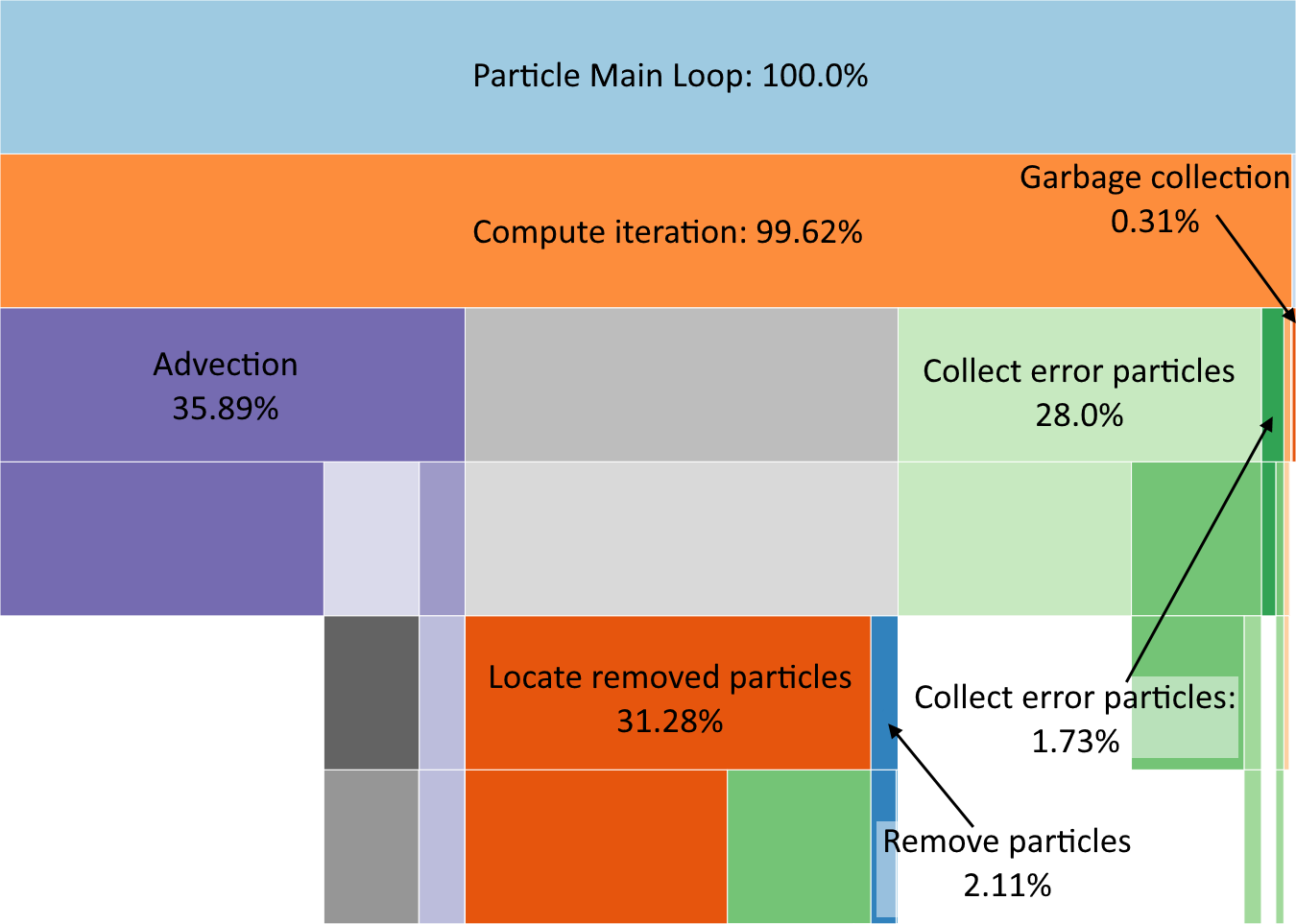}
			\label{fig:benchmarks:a}}
			\subfigure[dynamic particle removal]
			{\includegraphics[keepaspectratio, width=0.485\columnwidth]{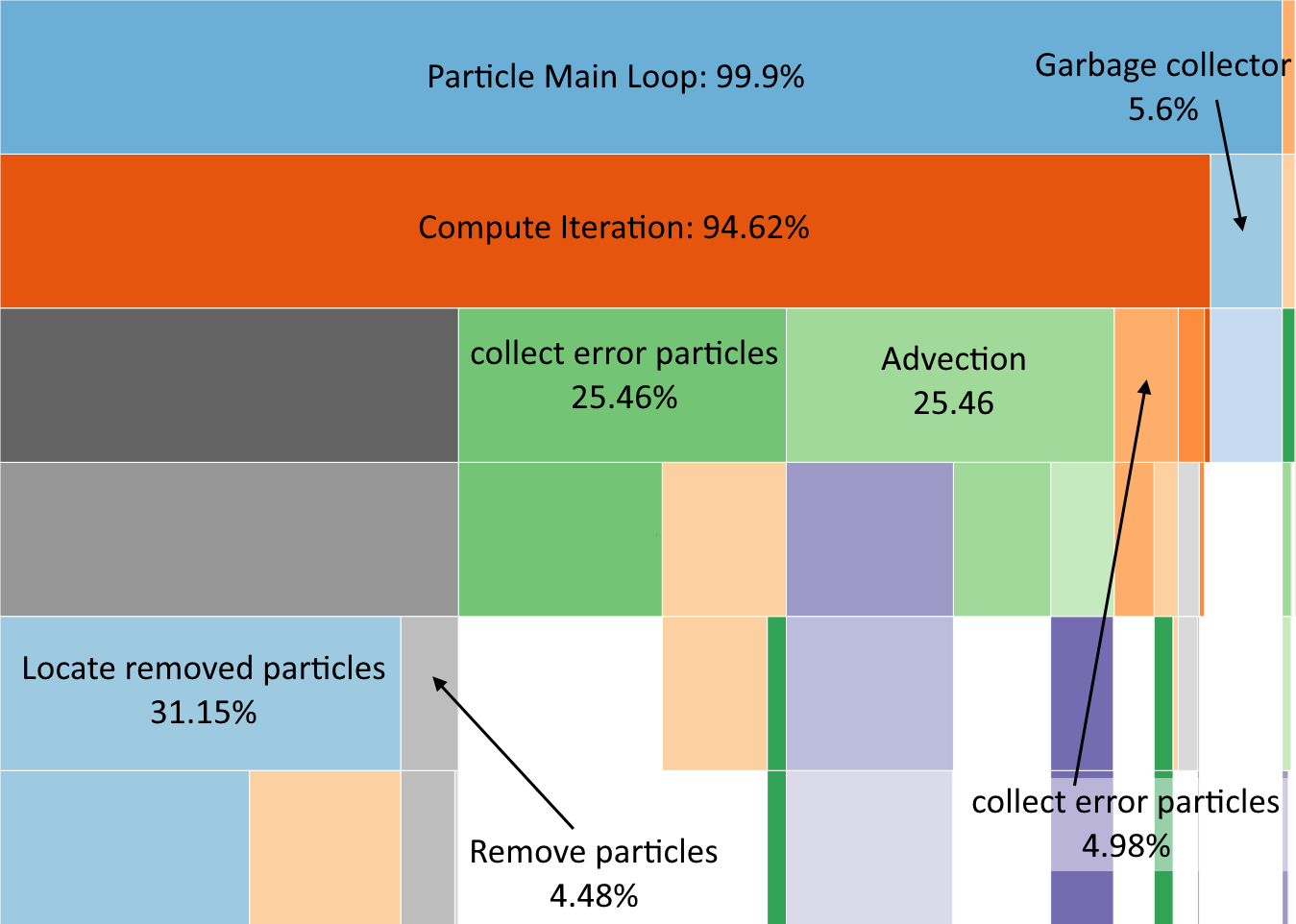}
			\label{fig:benchmarks:b}}
	 	\end{minipage}
    \end{center}
    \caption{Plot of per-codeline runtime distributions for different dynamic array operations. The average particle number target in both cases is $2^{18}$. Those plots show runs with no MPI-parallelisation and C-evaluated kernels.}
    \label{fig:benchmarks}
\end{figure}

The investigation into performance enhancing of Lagrangian particle simulations starts with the following research questions: \textit{What are the most costly functions and functionalities within the existing particle simulation?} In order to answer this question, we needed to (i) determine the most costly functions within the simulation framework, (ii) quantify the controls (i.e. impact parameters) on the computational runtime, and (iii) test the different computational settings of the simulation, such as Python- and C-based kernel evaluation, the impact of task parallelisation via message passing (MPI), and the cost of dynamic array operations. For this quantification, the Parcels code\footnote{OceanParcels website - \url{http://oceanparcels.org/}} in version 2.1.5 was profiled using the \textit{cProfile} package. The outcome were line-by-line code timings, which were analysed with \textit{snakevis}\cite{Davis2014}, plotting a percentage runtime as in \cref{fig:benchmarks}.

\begin{figure}[htbp]
  \centering
  \includegraphics[keepaspectratio, width=0.98\columnwidth]{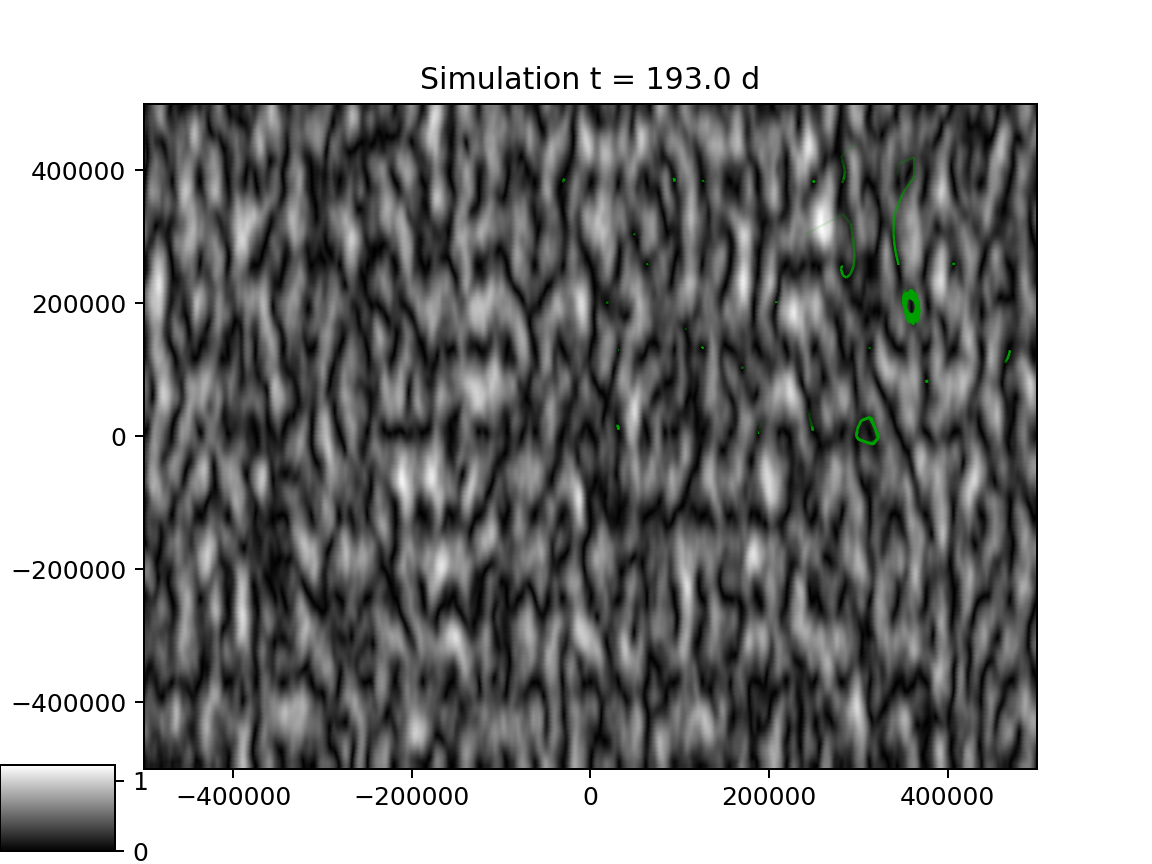}
  \caption{Probabilistic, synthetic fluid flow field, generated from perlin noise \cite{Perlin1985}. The background displays absolute flow speeds in range \{0, ..., 1.0\} $\frac{m}{s}$ in a metre-scale flow field, whereas the green particle traces depict the particles' trajectory within the flow.}
  \label{fig:perlin_flow}
\end{figure}

\Cref{fig:benchmarks} shows the total runtime (blue bars) of a 365-day particle simulation on a synthetic flow field generated by perlin noise (see \cref{fig:perlin_flow}), advected with a fixed temporal resolution of 1 hour (13,000 seconds for the removal-scenario, 265,000 seconds for the insertion-scenario). In both presented cases, the simulation targets an average number of particles in the \textit{array}-collection of $2^{18}$. At the third code level, one can see a first significant difference: the active garbage collector (darker green bar) consumes a significant amount of the runtime (5.6\%) for dynamic removal, which is not the case for the dynamic-insertion scenario. Furthermore, in the third and fifth level we see that both scenarios spend large amounts of time in localising erroneous (30.46\% for removal-scenario; 29.73\% for insertion-scenario) and deleted (31.15\% for removal-scenario; 31.28\% for insertion-scenario) particles. The removal scenario, where particles are actually deleted, also spends approximately twice the time in the removal function than the insertion scenario (4.48\%, compared to 2.11\%). Those extra runtime expenses for item removal and garbage collection result in less time being spent in the actual particle advection (only 25.46\% for the removal scenario, compared to the 35.89\% for the insertion-scenario). The remaining percentages are smaller helper functions, list-comprehensions and memory management. We note here again that the synthetic flow field on which particles are advected is held entirely in-memory and that particle positions are not written to external storage, so that extensive file-I/O is not a limit in those cases.

The following top-5 most expensive execution functions emerge from an extensive benchmark analysis of the different scenarios described above (i.e. static-vs.-dynamic datasets, varying particle set sizes, etc.). Furthermore, these are the most-expensive functions as their detrimental impact on performance scales with the size of the particle set, and hence are over-pronounced even when running large particle sets in parallel to reduce the common file-I/O performance malus in real-world ocean simulations.

\begin{enumerate}
    \item list comprehension in kernel execution that gathers all unadvected particles (computational complexity: $\mathcal{O(N)}$)
    \item \textit{delete()} function for removing individual particles from the particle set (computational complexity: $\mathcal{O}(\mathcal{N}^2)$)
    \item array-copy operation that flips the dense-array arrangement in memory from column-major order (FORTRAN/Python-native orientation) to row-major order (C/C$^{++}$-native orientation) for C-internal kernel evaluation (computational complexity: $\mathcal{O(N)}$)
    \item the actual particle advection computation function, which would ideally be at the top of this list (34\% of the runtime)
    \item another array transpose-and-copy operation that performs the nD-dense-array concatenation for periodically-released particles (computational complexity: $\mathcal{O}(2 \mathcal{N})$)
\end{enumerate}

Therefore, the clear target of this research is to reduce the impact of the above-listed (super-)linearly runtime scaling of collection re-ordering- and alteration functions, to the point that the advection function becomes the major runtime expenditure of the simulation. This would, as traditionally intended, transform the whole simulation process into a compute-bound application, which can subsequently be sped up by common high-performance computing approaches, such as SIMD processing, parallelisation and load distribution.

The reason for the runtime malus in the above-listed operations is that dense-arrays are random-access data structures that exhibit a high computational- and memory complexity for random alterations. In other words, adding and removing entities in an \textit{array}-collection requires splitting the collection in multiple distinct subsets that exclude removed- and include inserted elements, and concatenate them together into a newly-allocated result collection (i.e. they are not \textit{in-place} operations). Thus, we need to exploit alternative data structures that exhibit a sub-linear scaling on those operations.

\section{High-performance \textit{collection} data structures for Python-based numerical simulations}
\label{sec:method}

In this paper, we propose to supplement Python with dynamic collections, such as double-linked node-lists, bisect-search augmented arrays, with explicit C-bindings to perform high-speed kernel executions in C. Those structures attempt to remedy the performance malus rooted in slow random-alteration methods on \textit{array}-collections.

\begin{figure}[htbp]
  \centering
  \includegraphics[keepaspectratio, width=0.8\columnwidth]{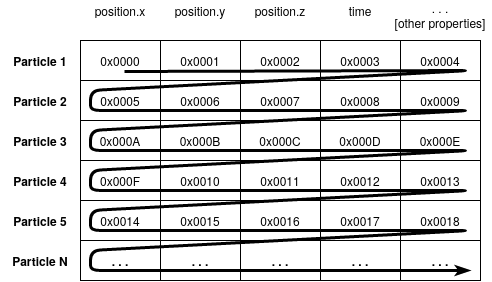}
  \caption{Array-of-Structure (AoS) collection memory layout - a collection data structure with an item-linearized memory alignment. The linear item order is annotated by the black major arrow.}
  \label{fig:method:aos}
\end{figure}

Our comparison baseline is a common Array-of-Structure (AoS) \textit{array}-collection data structure, which is a simple array housing complex multi-dimensional objects (see \cref{fig:method:aos} and the supplement material for details). The simplicity is advantageous for the conversion of the data structure into C, but it suffers from the severe speed malus for dynamic alterations.

In order to address the trivial and most common request to do engineering optimisations, such as speedy list comprehensions, reduced memory copies and so forth, we created a first comparison case of \textit{engineering-optimised AoS}, to showcase the potential as well as the limitations of simplistic and trivial improvements.

\subsection{List-of-Arrays [-of-Structures] (LoA[oS]) collection}
\label{sec:method:loa}

A first attempt to provide an upper bound to the scaling complexity of random array alterations is to split up the global array into a list of fixed-length arrays. We refer to this data structure as \textit{List-of-Array (LoA)} collection in the text remainder, though highlighting that the underlying array structure is \textit{AoS} by itself.

\begin{figure}[htbp]
  \centering
  \includegraphics[keepaspectratio, width=0.98\columnwidth]{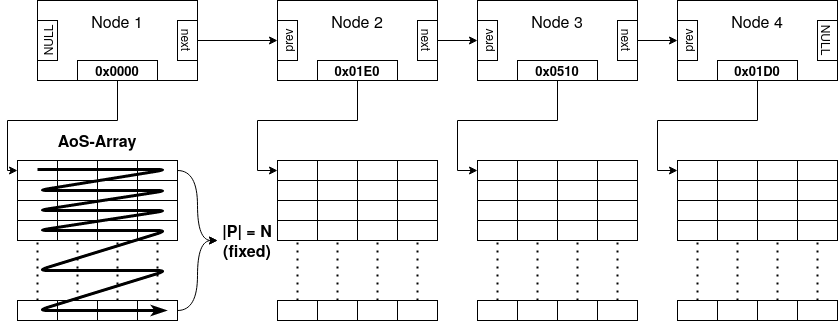}
  \caption{List-of-Array (LoA) collection memory layout - a node-list container with fixed-size, pre-allocated, \textit{AoS}-organised sub-arrays.}
  \label{fig:method:loa}
\end{figure}

Such a splitting setup subsequently requires an address translation when randomly accessing and retrieving items of the collection, hence slowing down excessively-numerous item look-ups. The advantage is that the complexity of item insertion- and deletion is reduced from $\mathcal{O(N)} = K * (N + (N \pm 1) + (N \pm 1) )$, with $K$ being the number of added or removed items and $N$ being the total number of items (i.e. particles), to $\mathcal{O(N)} = K * ( D + M + (M \pm 1) + (M \pm 1) + (D + 2\frac{M}{2})) $, with $D = \frac{N}{M}$ and $M$ being the maximum size of the sub-arrays and a loose bound that $N >> M$. The latter additional accounts for the effort of merging two sub-arrays when their occupancy is below 50\%. The layout is illustrated in \cref{fig:method:loa}.

\subsection{Structure-of-Arrays (SoA) collection}
\label{sec:method:soa}

The SoA representation is a transposition of the memory layout of AoS. Rather than using a single instance of the container datatype at the top-level, whose elements are particle structures, there is only one instance of the particle structure with an array of values in each field (e.g. velocity, position). A particle is then denoted by an index, and can be reconstructed by extracting the element at that index from all the constituent fields.

\begin{figure}[htbp]
  \centering
  \includegraphics[keepaspectratio, width=0.9\columnwidth]{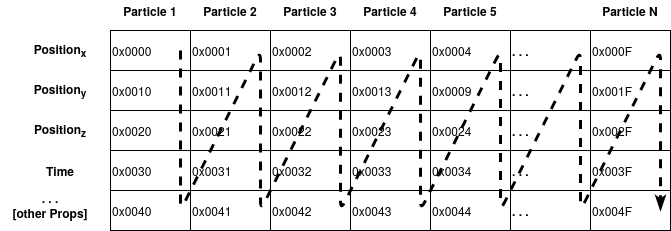}
  \caption{Structure-of-Array (SoA) collection memory layout - a collection data structure with an attribute-linearized memory alignment.}
  \label{fig:method:soa}
\end{figure}

Due to its layout, SoA lends itself to a vectorized instruction flow, as common to modern linear-algebra software frameworks (i.e. BLAS, LAPACK, NumPy) \cite{Intel2014}. When accessed from Python, a loop over all particles may be turned into a loop of NumPy-accelerated vectorized operations over the particle fields. This is particularly apparent in the case of multiple particle initialization, where the allocation of many independent objects can be replaced by the instantiation of just a few arrays. SoA may also be more cache-friendly for the just-in-time advection calculation. A field value for multiple particles may be loaded into a single cache line (see \cref{fig:method:soa}), thus an optimizing compiler may choose to use vectorized CPU instructions to operate on multiple particles at once (see \textit{SM1} in the supplementary material).  

The linearized in-memory layout of SoA is a trade-off of speed for dynamic flexibility. Cache coherence and high data locality in vectorized processing are prioritized over efficient deletion of arbitrary particles. Accessing and manipulating individual particles also becomes unwieldy, as all the fields are accessed separately.

%
%

\subsection{\textit{Nodal} Double-Linked List collection}
\label{sec:method:nodes}

The proposed \textit{LoA}-collection enforces an upper bound on the computational complexity of dynamic collection alterations, while the \textit{SoA}-collection inherently only improved cache access during advection execution, hence not underpinning tackling the dynamic collection alteration issue. In order to reduce that computational complexity, the actual dynamic collection alteration process needs to be addressed.

We here introduce a double-linked list implementation into the Parcels particle framework. The novel aspect here is its straightforward implementation with \textit{ctypes}, which allows direct access of the data collection within the high-speed  C-kernels.

\begin{figure}[htbp]
  \centering
  \includegraphics[keepaspectratio, width=0.9\columnwidth]{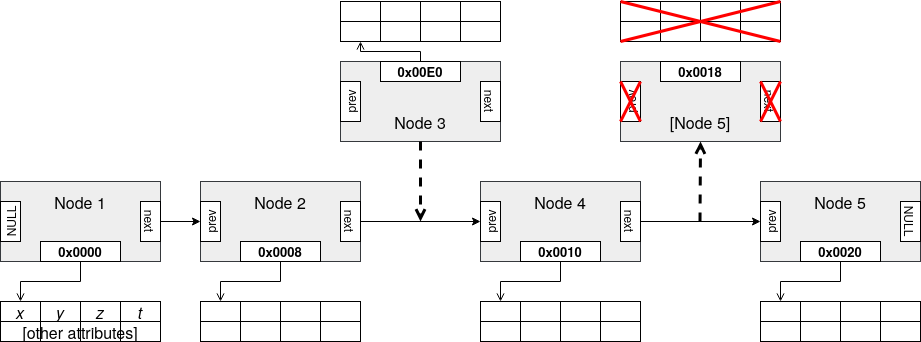}
  \caption{Node-based double-linked list collection memory layout.}
  \label{fig:method:nodes}
\end{figure}

In this implementation, the Node objects are a complementary of Python class-objects and their C-representation as a structure. Each node stores a reference to its predecessor, its successor and its data container. The data container in this case is a particle-object reference (in Python) and its NumPy-object pointer reference (in C). This NumPy-object reference of a particle is just a pointer to a NumPy-mapped C-struct reference of one particle, not a pointer to a NumPy array.

During runtime, particles can dynamically be added and removed with the standard complexity of $\mathcal{O(1)}$ for linked lists. Particles are bound to a Node, and insertion and removal is performed by plain relinking of the predecessor- and successor references in the collection. In contrast to Python's internal \textit{deque} implementation, the list can commonly be traversed by begin-end-iterators or foreach-element-iterations in Python and C equivalently. The traditional drawback remaining here too is the lack of (fast) random-access, which usually requires an $\mathcal{O(N)}$ list traversal compared to the $\mathcal{O(1)}$ random-access in arrays. Thus, it is expected that the nodes-based double-linked list outperforms alternative implementations in scenarios with abundant dynamic particle operations.

For abbreviation purposes, the remainder of the paper refers to the double-linked list as \textit{nodal list}, as our benchmark logs- and plots refer to the collection as \textit{nodes}. In the context of this paper, it is hence implied that nodes in a \textit{nodal list} are double-linked.

\section{Results}
\label{sec:results}

In order to assess, compare and quantify the effectiveness of the individual data collection designs, we run benchmarks on the same setup as in \cref{sec:statusquo}. In contrast to the initial benchmark, timings do not need to be resolved down to individual atomic operations, which costs considerable runtime and memory in and of itself. For evaluation purposes, the parameters of interest are listed as follows, and measured at the highest distinct code level:

\begin{itemize}
    \item \textit{Compute time}: the bare time required to execute flow advection kernels and field interpolations
    \item \textit{I/O time}: the combined time require for \textit{file-I/O} and \textit{memory I/O} operations
    \item \textit{Plotting time}: operation time required for console- and image output (i.e. image rendering)
    \item \textit{Total time}: the time between the start of the simulation and the end of the simulation, excluding: pre-processing, setup of flow-fields and initial collections, start-up phase, result collection and write-back, clean-up, post-processing and shutdown. Approximately equal to the cumulative compute-, I/O- and plotting time, with minor round-up deviations possible.
    \item File-I/O time: first component of I/O-time that measures all file-related operation times (i.e. data transfer between external storage and memory); the demonstrated benchmarks also include memory-I/O of the gridded flow fields in this measure, because the use of synthetic flow fields is a special case and the interface commonly accesses files instead of pre-loaded memory sections.
    \item Memory-I/O: second component if I/O-time that measures memory management operation times (e.g. array- and variable copies; memory mapping between C- and Python storage).
    \item Kernel-times: average \textit{compute}-, \textit{I/O}- and \textit{total} processing time among all iterations of the simulation.
    \item Per-particle time: average \textit{compute}-, \textit{I/O}- and \textit{total} processing time for an individual particle per simulation iteration.
    \item \textit{Memory consumption}: the amount of memory acquired by the simulation process for instruction-, data- and stack segment, as well as the size of shared memory page blocks.
\end{itemize}

Furthermore, to enable appropriate comparisons, static and dynamic scenarios are setup to cover in each simulation approximately the same average number of particles, and hence the same amount of workload during the simulation time, while avoiding irrelevant setups (e.g. simulations with $\leq 4$ particles, single-particle additions and removals each iteration, etc.). In order to achieve that with dynamic scenarios, we start with $N = \frac{N_{target}}{2}$ items and insert 128 items with a periodicity $t_p = (t_{sim} / \frac{N}{128}) / \frac{7}{4}^2$, while removing items by defining a randomized, uniformly-distributed life expectancy $p(x) = \frac{1}{b-a}$ with $b = 2 \sqrt{\frac{2}{3}} (t_{sim} - t)$ and $a = 0$.

In order to cover a wide range of scenarios, we benchmark each experiment with $N = 2^{x}$ particles, with $x \in \lbrace 10 ... 16 \rbrace$.

\subsection{Static Datasets}
\label{sec:results:const}

\begin{figure}[htbp]
	\begin{center}
	    \includegraphics[keepaspectratio, width=0.6\columnwidth]{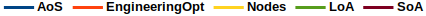}
	 	\begin{minipage}{\columnwidth}
        \centering
			\subfigure[Total time]
			{\includegraphics[keepaspectratio, width=0.319\columnwidth]{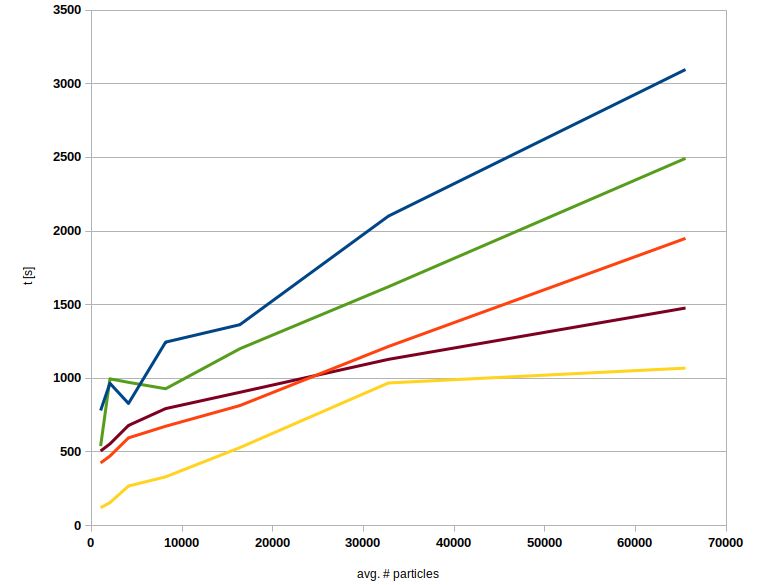}
			\label{fig:results:static:top:a}}
			\subfigure[Average runtime per particle]
			{\includegraphics[keepaspectratio, width=0.319\columnwidth]{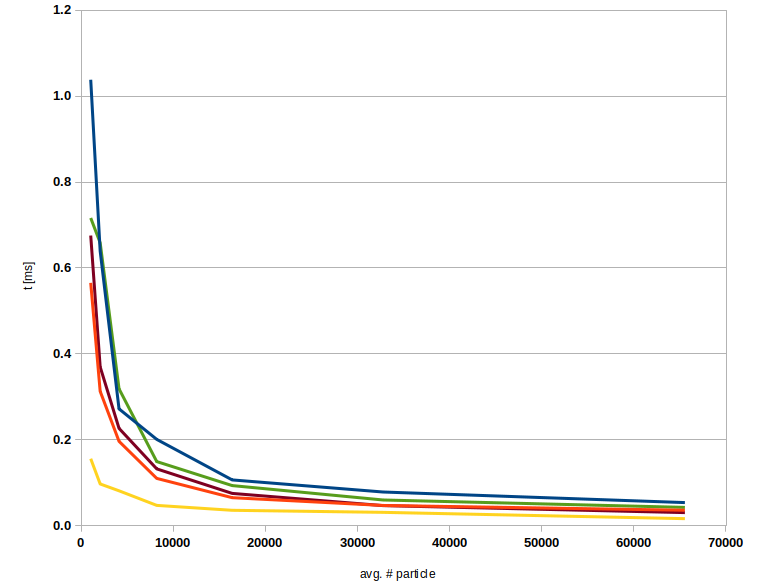}
			\label{fig:results:static:top:c}}
			\subfigure[Memory consumption]
			{\includegraphics[keepaspectratio, width=0.319\columnwidth]{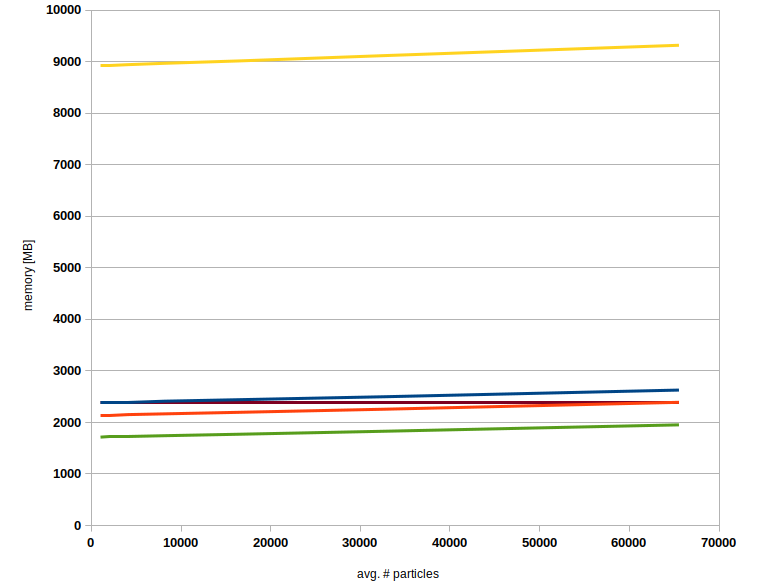}
			\label{fig:results:static:top:b}}
	 	\end{minipage}
    \end{center}
    \caption{Runtime behaviour of different collection data structures for a static simulation setup.}
    \label{fig:results:static:top}
\end{figure}

A static setup with a fixed particle number within the collections shows the runtime behaviour in \cref{fig:results:static:top}.

All adapted collection types consistently improve upon the runtime when compared to the \textit{AoS}-case. However, we observe rather diverging runtime behaviours: \textit{LoA} and the engineering case show a linear time increase, which is just negatively offset to the \textit{AoS}-curve. This is rather unsurprising as both collection structures are conceptually or structurally similar to the \textit{AoS} starting case. The \textit{SoA} structure shows a strong logarithmic time behaviour whereas the double-linked list shows a weak logarithmic behaviour. Both of those behaviours agree with the approximations of the computational complexity for data access (see \cref{sec:method:nodes}). 

The overall consistently fastest collection implementation is the nodal list. This statement is not only supported by the overall runtime (\cref{fig:results:static:top:a}) but even more so by the average runtime per individual particle (\cref{fig:results:static:top:b}). The plot shows the limited constant offset in time for setting up and managing the data structure, as the per-particle runtime even for small datasets is significantly lower than the other structures. This beneficial runtime behaviour comes, however, with a significantly-increased memory demand, as each item record (i.e. each node with its particle data package) needs to store the predecessor and successor reference.

\begin{figure}[htbp]
	\begin{center}
	    \includegraphics[keepaspectratio, width=0.6\columnwidth]{legend}
	 	\begin{minipage}{\columnwidth}
        \centering
			\subfigure[Compute vs. I/O ratio]
			{\includegraphics[keepaspectratio, width=0.319\columnwidth]{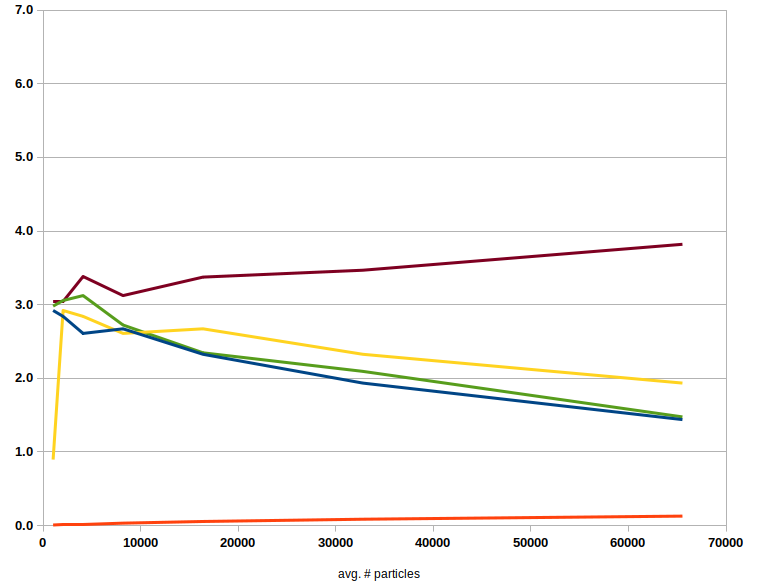}
			\label{fig:results:static:bottom:a}}
			\subfigure[Speed-Up (relative to \textit{AoS})]
			{\includegraphics[keepaspectratio, width=0.319\columnwidth]{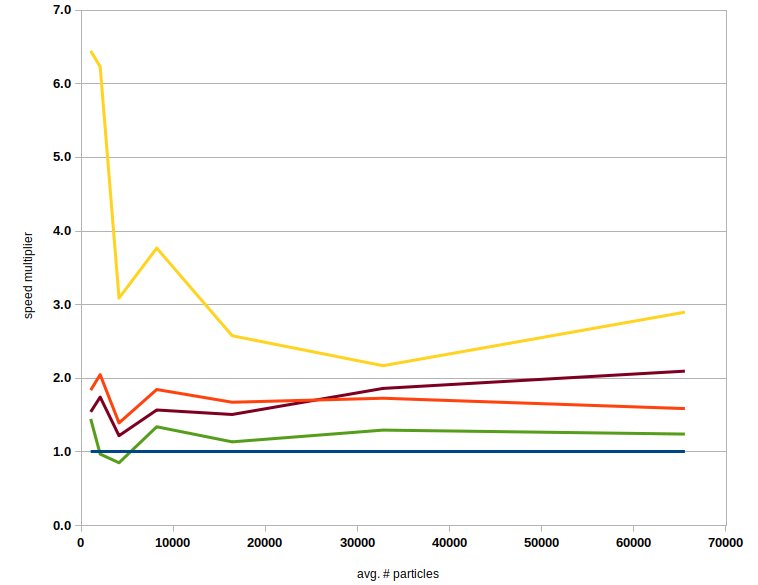}
			\label{fig:results:static:bottom:b}}
	 	\end{minipage}
    \end{center}
    \caption{Derivative performance metrics of different collection data structures for a static simulation setup.}
    \label{fig:results:static:bottom}
\end{figure}

In \cref{fig:results:static:bottom:b}, we see that the speed-ups for minor data structure modifications compared to \textit{AoS} (e.g. \textit{LoA} and engineered improvements) is limited, which is at no point exceeding the 2x speed-up. Different collection structures, such as \textit{SoA} and nodal lists, yield higher speed-ups. \Cref{fig:results:static:bottom:b} also shows that \textit{SoA}'s starts with less speed-up for small collections and a consistently increasing speed-up for larger collections. In contrast, the speed-up of nodal lists is highest with smaller collections, then depreciates for intermediate-sized collections while then increasing again for very large datasets. The runtime (and hence also speed-up) is less reliable and stable for nodal lists than for \textit{SoA} collections. 

Furthermore, \cref{fig:results:static:bottom:a} shows that, for most collections, more time is spent in memory management tasks the larger the dataset becomes, meaning that the simulation process is increasingly I/O-bound - more specifically: memory-I/O bound. The single exception to this behaviour is the \textit{SoA} data structure, which inherently requires minimal memory management when static dataset sizes are concerned, leading to an increasing compute-to-I/O ratio. 

\subsection{Fully-Dynamic Datasets}
\label{sec:results:addage}

\begin{figure}[htbp]
	\begin{center}
	    \includegraphics[keepaspectratio, width=0.6\columnwidth]{legend}
	 	\begin{minipage}{\columnwidth}
        \centering
			\subfigure[Total time]
			{\includegraphics[keepaspectratio, width=0.319\columnwidth]{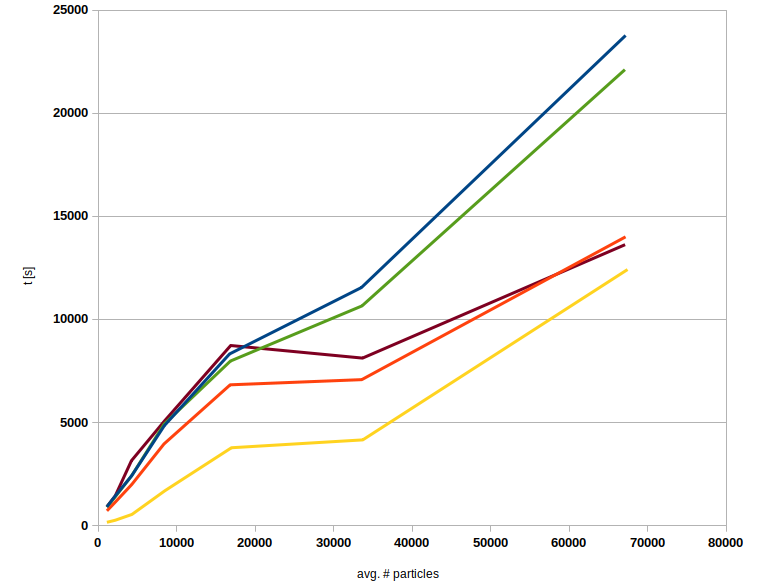}
			\label{fig:results:dynamic:top:a}}
			\subfigure[Average runtime per particle]
			{\includegraphics[keepaspectratio, width=0.319\columnwidth]{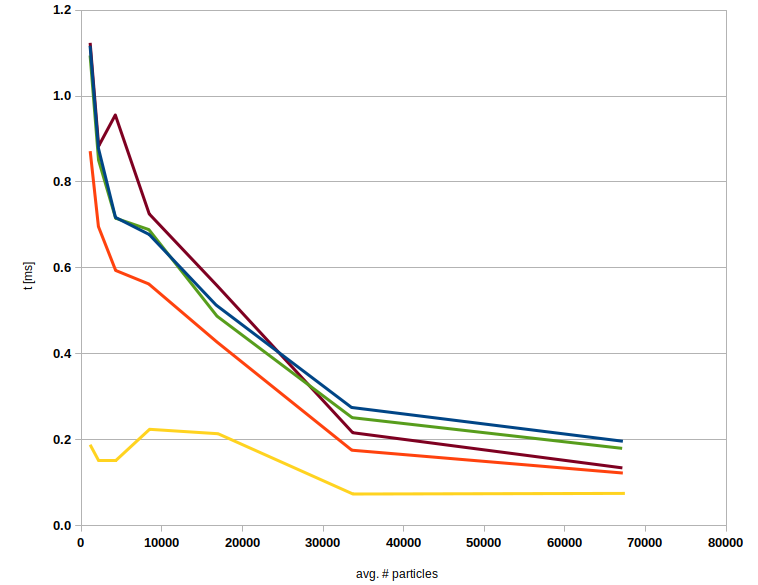}
			\label{fig:results:dynamic:top:c}}
			\subfigure[Memory consumption]
			{\includegraphics[keepaspectratio, width=0.319\columnwidth]{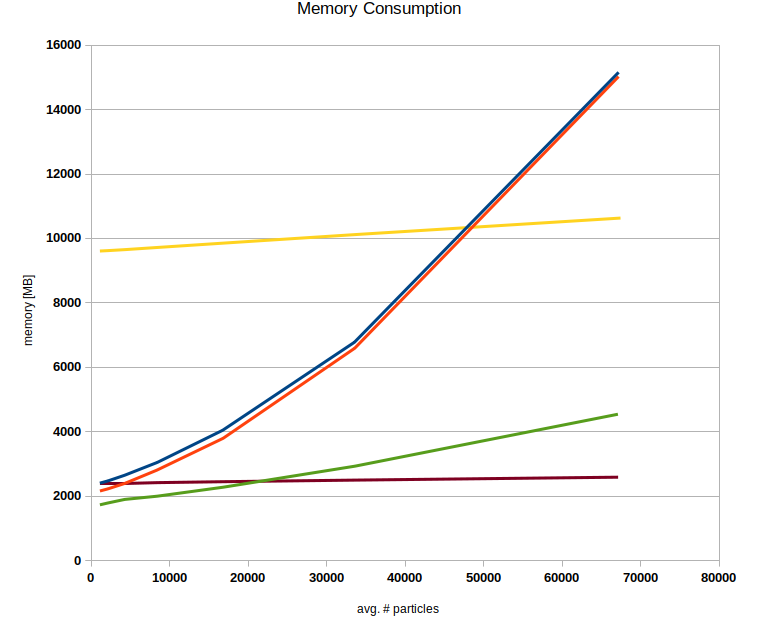}
			\label{fig:results:dynamic:top:b}}
	 	\end{minipage}
    \end{center}
    \caption{Runtime behaviour of different collection data structures for fully dynamic collection operations during simulation.}
    \label{fig:results:dynamic:top}
\end{figure}

In the dynamic case, when combining the dynamic collection operations of insertion and removal, the performance figures significantly change, as seen in \cref{fig:results:dynamic:top}.

\begin{figure}[htbp]
	\begin{center}
	    \includegraphics[keepaspectratio, width=0.6\columnwidth]{legend}
	 	\begin{minipage}{\columnwidth}
        \centering
			\subfigure[Compute vs. I/O ratio]
			{\includegraphics[keepaspectratio, width=0.319\columnwidth]{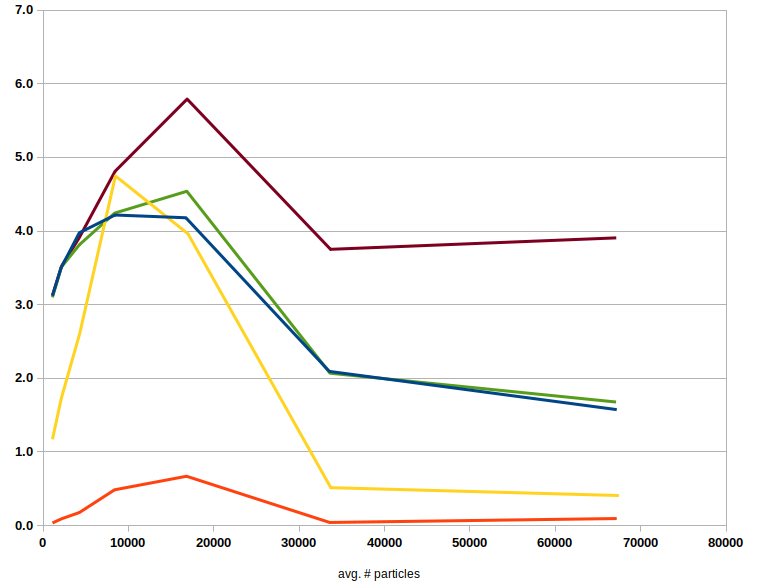}
			\label{fig:results:dynamic:bottom:a}}
			\subfigure[Speed-Up (relative to \textit{AoS})]
			{\includegraphics[keepaspectratio, width=0.319\columnwidth]{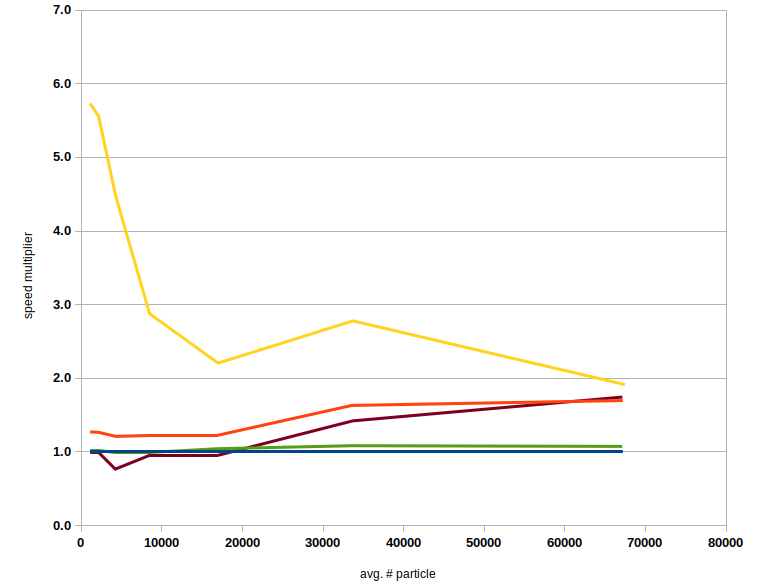}
			\label{fig:results:dynamic:bottom:b}}
	 	\end{minipage}
    \end{center}
    \caption{Derivative performance metrics of different collection data structures for fully dynamic collection operations during simulation.}
    \label{fig:results:dynamic:bottom}
\end{figure}

Here, we first separate the analysis in two regimes: small collections (i.e. low particle count) with $N \leq 25,000$ and large collections with $N > 25,000$. In small collections, we see two speed regimes, namely the array-like structures (i.e. \textit{AoS}, engineering improvements, \textit{LoA} and \textit{SoA}) and the double-linked list. Array-like structures experience a hyperbolic rise in runtime for small collections, because computational effort still outweighs the additional memory-I/O workload (see also \cref{fig:results:dynamic:bottom:a}). Furthermore, the fixed memory allocation requires more setup time than the dynamics of a nodal lists, which clearly outperforms the other collections.

In the second regime of large collections, we see further divergence of the data structures into three categories: (Category A) row major order-aligned array-like collections (i.e. \textit{AoS} and \textit{LoA}), (Category B) column major order-aligned array-like collections (i.e. engineering improvements and \textit{SoA}) and (Category C) the nodal list. For large collections, Cat. A collections scale linearly and with a steep slope relative to $N$, due to their disadvantageous memory layout, which complicates memory operations via NumPy. Cat. B collections exhibit a more logarithmic runtime scale for large collections. That is because, despite their fixed memory arrangements, NumPy internally concatenates column major order arrays faster than the row major order arrays. For array-like data structures in terms of runtime, both insertions and removals result in large number of array concatenation operations. As expected, the double-linked list more favourably deals with dynamic collection operations. When comparing the results in \cref{fig:appendix:insertion} and \cref{sec:results:addage}, double-linked lists are the only collections structures that exhibit approximately equal runtimes for both cases. The regime split as well as the performance argumentation is further supported by the compute-to-I/O ratio in \cref{fig:results:dynamic:bottom:a}, as well as an observation of the speed-up in \cref{fig:results:dynamic:bottom:b}.

In comparison to previous experiments, the \textit{SoA} structure seems particularly impeded in performance from random operations, as it is consistently slower than the engineering improvements. Furthermore, the specific advantages and disadvantages balance the performance of \textit{AoS}, \textit{LoA} and \textit{SoA} for intermediate dataset sizes. For very large collections, the per-particle runtime is approximately equal for nodal lists, engineering improvements and \textit{SoA} collections.

The final point of note, and also the most interesting observation for extrapolating the results to massive dataset sizes with $N >> 100,000$, are the observations of memory consumption. The fully dynamic operations and their subsequent concatenation operations leave temporary data blocks as traces in memory, leading to an exponential memory demand for row major order-aligned array-like collections (Cat. A collections). The fixed-size buckets of \textit{LoA} prevent such an excess in memory consumption. For \textit{SoA} and nodal lists, their memory requirements remain linear in our measurements.

Hence, combining the insights in runtime, per-particle runtime and memory behaviour, the only two viable collection data structures for massive, dynamically-altering datasets (i.e. huge particle swarms in the fluid simulation) are double-linked (nodal) lists and \textit{SoA}. The double-linked list yields the highest performance in terms of speed, but requires five times more memory than \textit{SoA} due to the supplemental node information. If memory capacity is the major limiting factor, then the \textit{SoA} data structure delivers the maximum speed for the least memory demands.

\section{Discussion}
\label{sec:discussion}

The runtime- and memory consumption results above demonstrate the diverse performance impacts that the choice of a specific collection data structure has on large-scale simulations. Traditionally, the choice of the collection data structure depended exclusively on

\begin{itemize}
    \item the dataset size
    \item the available compute-related technologies (e.g. SIMD, parallelization, assembly compilers)
    \item the available memory
\end{itemize}

Our benchmark analysis demonstrates that the choice of the ideal collection data structure depends on the scenario and the operations actually performed on the dataset. 

In order to judge potential future improvements, we analyse the compute-to-I/O ratio. Here, while the static- and dynamic case appear very different, their trends on large data are similar: all new collection types show an increased computational load for $N \leq 25,000$ items. Larger collections experience a drop-off in the ratio. On very large collections with $N \geq 50,000$ items, only the \textit{SoA} collection experiences another increase in compute load. In conclusion, all collections with a bad data locality show an I/O-bound processing behaviour, whereas the cache-optimised \textit{SoA} collection has a compute-bound processing behaviour. The data locality issue also explains the large drop-off in the compute-to-I/O ratio for nodal lists: items of the collection are sparsely distributed in the memory, in contrast to any array-like collection. This impacts the ideal use of the collections, as nodal list perform well with dynamic collection operations and simple compute kernels, such as the simple advection used in our experiments. The \textit{SoA} collection is expected to conversely outperform the other collections in static- and dynamic experiments complex computational kernels and an increasing amount of field interpolations. Furthermore, the high degree of data locality is beneficial to shared-memory parallelization. In the future, we shall combine the double-linked list collection with an improved data locality scheme.

Our findings directly impact studies in ocean physics, which can be split into the following case-study categories:

\begin{enumerate}
    \item \textit{Mass- / Momentum-conserving particle studies}: Lagrangian particles are once released and traced over stretched time periods. The set of particles stays fixed (i.e. \textit{static}).
    \item \textit{End-of-state particle studies}: particles are once released, their state is traced over multiple iterations, and they are removed upon reaching a given \textit{end state}. The simulation either ends after removal of the last particle, or releases particles anew when a particle converges on its end state. The simulation either just includes dynamic data removal, or is fully dynamic. A special case of such an end-of-state simulation is \textit{particle aging}, where particles are deleted after a defined maximum lifetime.
    \item \textit{Density- and flux approximations}: particles are continuously released and their position and attribute modalities are traced. At regular intervals, a regular-gridded map of those attributes is created by averaging the the particle attributes over all particles covered within a gridcell (i.e. stencil buffer-like operation). In order to facilitate a high particle density, particles are constantly created and inserted to the collection. In order to prevent accuracy errors in the accumulation, long-traced particles are regularly removed from the dataset. This is a traditional case of large dynamics in the collection data structure that contains the particles.
\end{enumerate}

According the our findings, each of those simulation cases shall employ its optimal collection structure. Qualitatively, the data structures enable developing new, dynamic operations that are driven by the particle behaviour itself (e.g. split- or merge of particles). Such particle-inherent dynamics are beyond the capabilities of geophysical Lagrangian oceanic-flow simulations until now, as they require (i) a versatile definition of particle behaviour and (ii) an efficient, memory-conserving, fast data structure that facilitates random insertions and removals. Specific example applications are the individual modelling of fractured particulate matter in the oceans, or the aggregation of smaller particulates into larger clusters.


The performance results on the different data structures extend further to computational cases outside physics simulations: General data analysis requires iterative attribute derivation and subsequent cost-function computation, as well as information condensation by removing or merging attributes. On large datasets, those data operations map to the different experiments for static datasets, dynamic data insertion, dynamic data removal, and fully dynamic operations in \cref{sec:results}. Therefore, (geophysical) data science applications and frameworks can benefit from improved data organisation. This is particularly important where computation speed is crucial, such as in emergency response systems and forecast systems that are connected to real-time sensor networks.

Regarding the conceptual- and development approach, we acknowledge that performance optimisation and especially performance measurement are hardly possible in Python. The challenge for measurement and improvement is that Python's native core (e.g. typeless codes, class hierarchy, interpreter- and introspection support) is too slow to measure runtimes and memory consumption below a certain resolution (empirically $\Delta t_{m} \geq 20 ms$). Thus, temporary resource allocations that may happen within NumPy or SciPy cannot be measured in Python because its measurement function calls are too slow. This can be observed in the numbers (see \cite{Kehl2021_data}) and the nearly-flat memory consumption graph of \textit{SoA}. Furthermore, we acknowledge that the manual provision of just-in-time C-compiler interfaces with \textit{ctypes} itself consumes resources (i.e. runtime and memory) that may either be hidden, reduced or removed with other C-binding implementations (e.g. Numba).

In order to check the results or apply the outlined structures to other modelling- and simulation problems, the program codes for the individual structures are available on github:

\begin{itemize}
    \item \textit{AoS}: \url{https://github.com/OceanParcels/parcels/tree/benchmarking}
    \item \textit{Engineering Optimizations}: \url{https://github.com/OceanParcels/parcels/tree/engineering_optim_trials}
    \item \textit{LoA}: \url{https://github.com/OceanParcels/parcels/tree/list_of_pset_array_trials}
    \item \textit{Double-linked list}: \url{https://github.com/OceanParcels/parcels/tree/sorted_pset_trials}
    \item \textit{SoA}: \url{https://github.com/OceanParcels/parcels/tree/soa_benchmark}
\end{itemize}


\section{Conclusion}
\label{sec:conclusion}

This paper aimed at quantifying the detrimental performance impact of using static, array-like collection data structures for dynamic collection operations (e.g. item insertion and removal) in geophysics (and especially oceanic) applications. Following the expected negative impact of detrimental data locality, new high-performance collection data structures within Python were designed and developed that also interfaces easily with C-kernels. That is because, in actual physics calculations, the performance gain from optimised data structures could vanish if calculations then require a slow, native Python processing. The whole process is specifically analysed for large datasets cases ($N > 2^{15}$), for which those data structure performance limits are increasingly impacting the overall simulation runtime.

We have demonstrated and quantified the negative performance impact in our \textit{status-quo experiments} (\cref{sec:statusquo}), listing the five most runtime-costly code operations that increasingly limit performance with growing datasets. Qualitatively, calls to the insertion- (i.e. \textit{concatenate}) and deletion (i.e. \textit{remove}) functions, as well as related array re-ordering functions, are prime performance barriers.

Consequently, the paper introduces new collection data structures to circumvent the performance bottlenecks, of which two collections are array-like structures (i.e. \textit{SoA}, \textit{LoA}) and one is a double-linked list structure. Our experiments showed that using those newly proposed data structures yield significant runtime improvements. The paper shows the two major experiment cases: a static dataset (a) and a fully dynamic dataset with insertion and removal (b). In all cases, the double-linked list is the fastest-computing data structure whereas \textit{SoA} is the most efficient concerning runtime improvement and memory consumption in combination.

The results and insights gained from the specific Lagrangian oceanic- and fluid flow simulation are generalisable to other problems in (geo)physics as well as general data science, as demonstrated in \cref{sec:discussion}. The developed data structures and the gained insight on performance facilitate new application scenarios and modelling cases in physical oceanography, which demonstrates the impact of our research.


\appendix
\section{Removal-only and insertion-only experiments}
\label{sec:appendix:add_age_experiments}

\begin{figure}[htbp]
	\begin{center}
	    \includegraphics[keepaspectratio, width=0.6\columnwidth]{legend}
	 	\begin{minipage}{\columnwidth}
        \centering
			\subfigure[Total time]
			{\includegraphics[keepaspectratio, width=0.23\columnwidth]{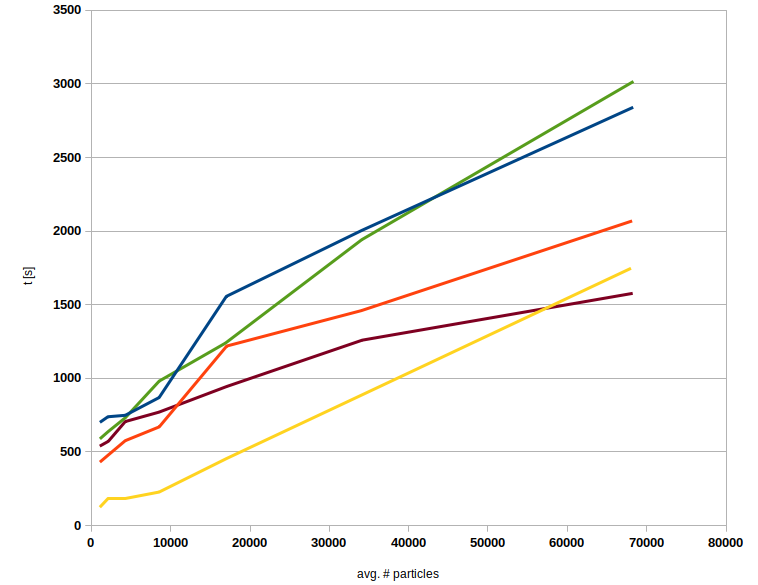}
			\label{fig:appendix:removal:a}}
			\subfigure[Memory consumption]
			{\includegraphics[keepaspectratio, width=0.23\columnwidth]{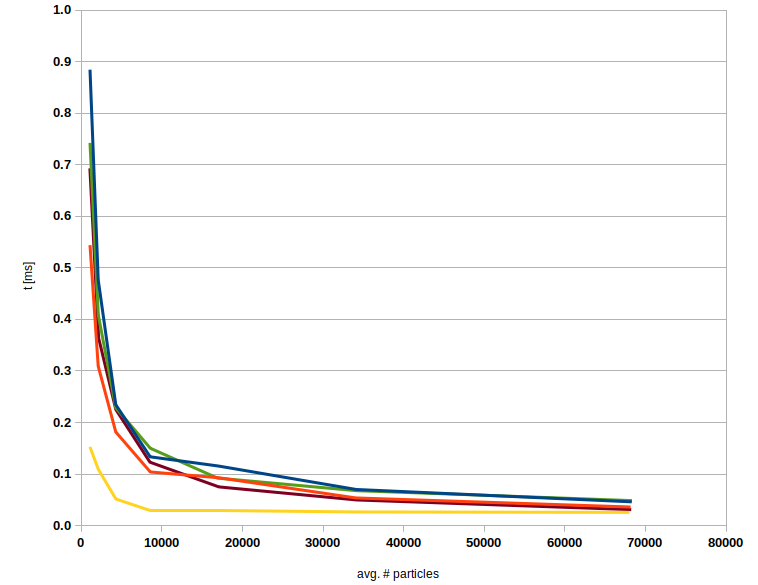}
			\label{fig:appendix:removal:b}}
			\subfigure[Compute vs. I/O ratio]
			{\includegraphics[keepaspectratio, width=0.23\columnwidth]{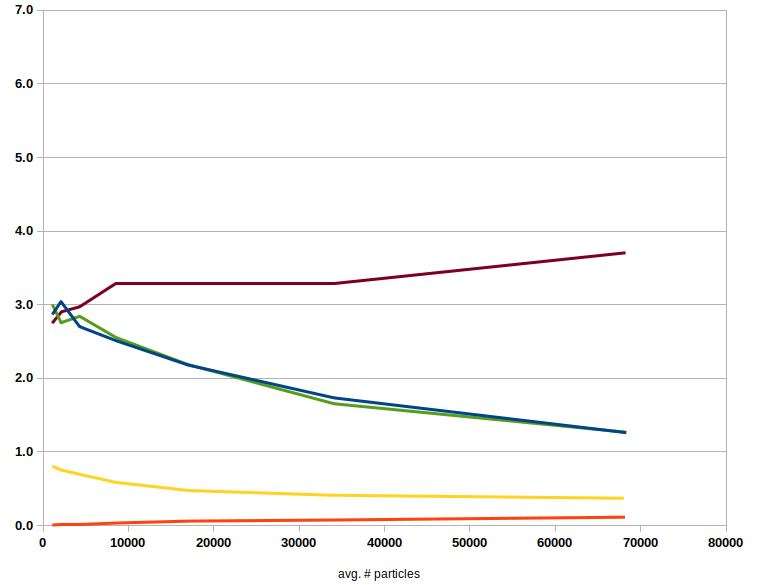}
			\label{fig:appendix:removal:c}}
			\subfigure[Speed-Up (relative to \textit{AoS})]
			{\includegraphics[keepaspectratio, width=0.23\columnwidth]{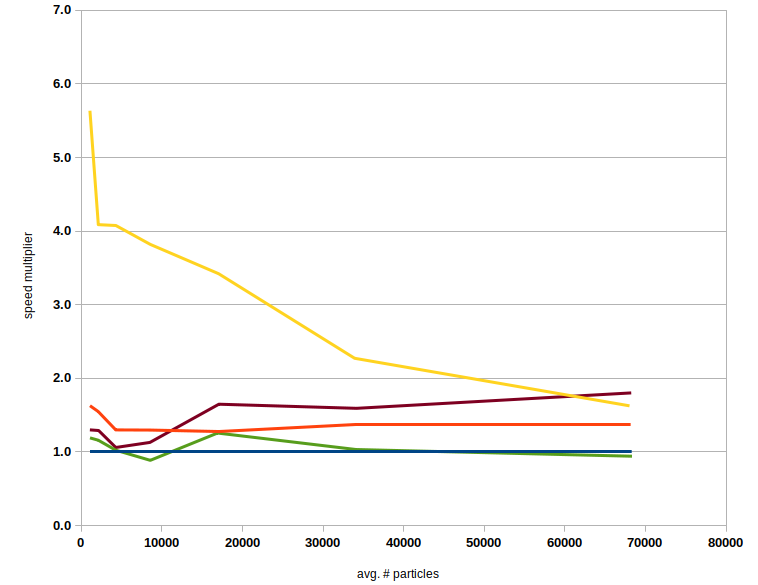}
			\label{fig:appendix:removal:d}}
	 	\end{minipage}
    \end{center}
    \caption{Runtime behaviour of different collection data structures for a dynamic removal of elements during simulation.}
    \label{fig:appendix:removal}
\end{figure}

\begin{figure}[htbp]
	\begin{center}
	    \includegraphics[keepaspectratio, width=0.6\columnwidth]{legend}
	 	\begin{minipage}{\columnwidth}
        \centering
			\subfigure[Total time]
			{\includegraphics[keepaspectratio, width=0.23\columnwidth]{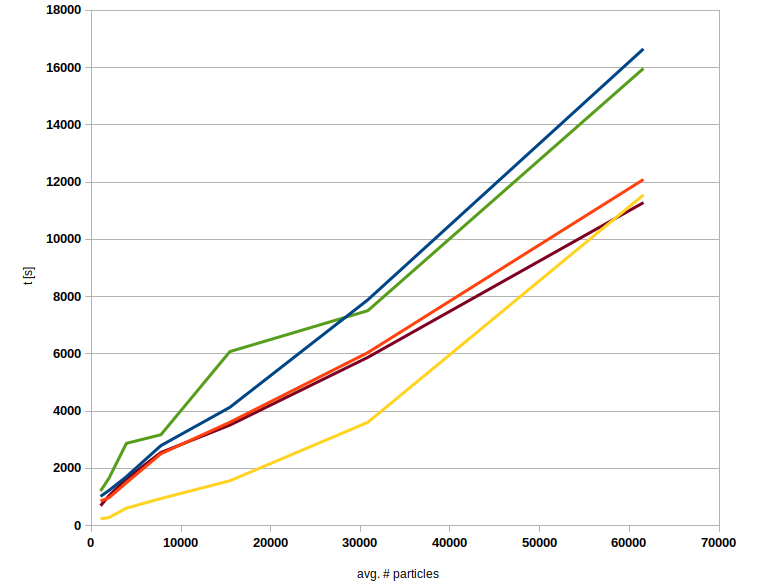}
			\label{fig:appendix:insertion:a}}
			\subfigure[Memory consumption]
			{\includegraphics[keepaspectratio, width=0.23\columnwidth]{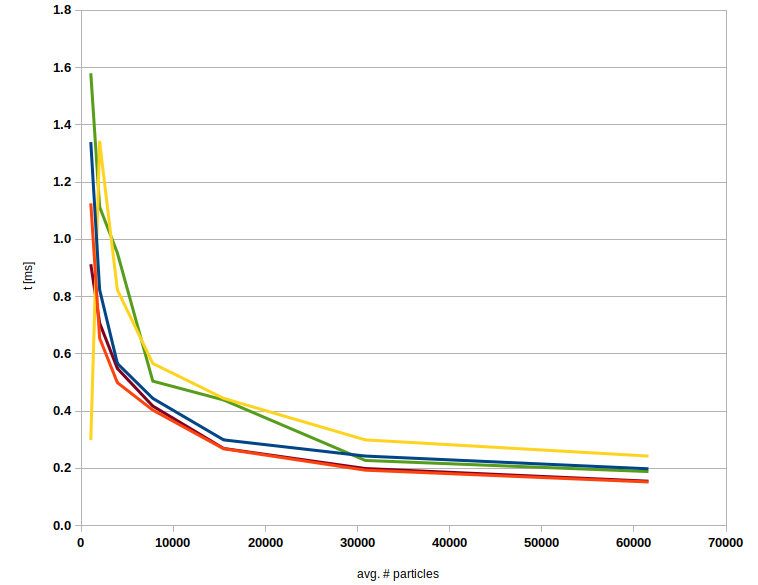}
			\label{fig:appendix:insertion:b}}
			\subfigure[Compute vs. I/O ratio]
			{\includegraphics[keepaspectratio, width=0.23\columnwidth]{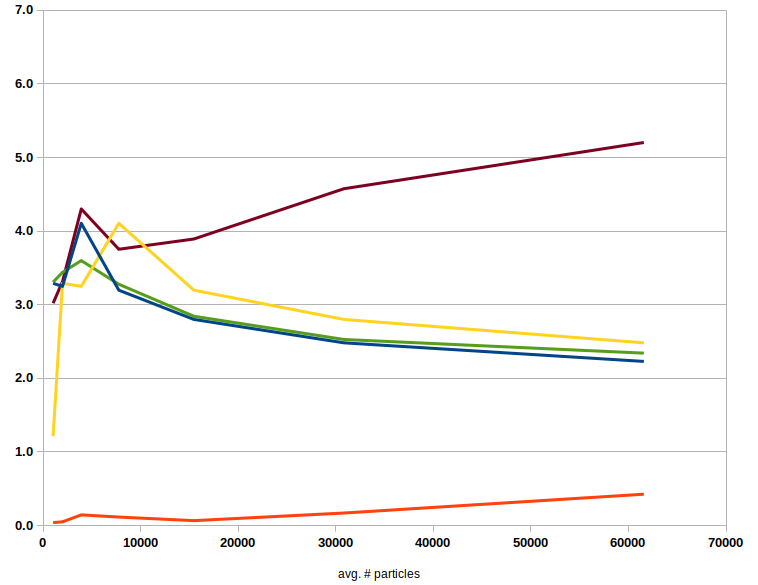}
			\label{fig:appendix:insertion:c}}
			\subfigure[Speed-Up (relative to \textit{AoS})]
			{\includegraphics[keepaspectratio, width=0.23\columnwidth]{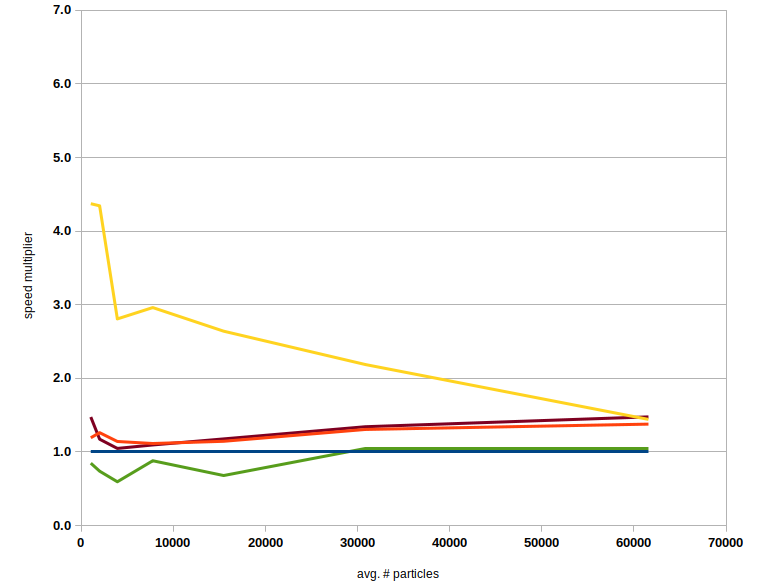}
			\label{fig:appendix:insertion:d}}
	 	\end{minipage}
    \end{center}
    \caption{Runtime behaviour of different collection data structures for a dynamic insertion of elements during simulation.}
    \label{fig:appendix:insertion}
\end{figure}

Separate experiments were conducted for removal-only (\cref{fig:appendix:removal}) and insertion-only(\cref{fig:appendix:insertion}) cases.

The removal-only experiment responds similar to the static case with two exceptions. Firstly, the \textit{LoA} collections is partially slower than the \textit{AoS} collection, which can be attributed to the \textit{merge}-operation of nodes with sublists of length $N \leq N_{lim}$. Secondly, the nodal list does not provide the expected speed-up compared to array-like collections, especially \textit{SoA}. This is because, with \textit{ctypes} as JIT-interface for the simulations, the node removal and re-linking requires additional operations to the traditional replacement of the \textit{next}- and \textit{previous} pointer.

The insertion-only case is similar to the fully-dynamic case, showing that particle insertion represents a major runtime expense. The distinct fast-insertion complexity of nodal lists is visible in its improved runtime. The \textit{SoA} performance is comparable to the nodal list because the particles are unordered, thus a particle insertion is performed by simple array concatenation.



\section*{Acknowledgments}
This work is part of the "Tracking Of Plastic In Our Seas" (TOPIOS) project, supported through funding from the European Research Council (ERC) under the European Union’s Horizon 2020 research and innovation programme (grant agreement no. 715386). Simulations were carried out on the Dutch National e-Infrastructure with the support of SURF Cooperative (project no. 16371 and 2019.034). 

\bibliographystyle{siamplain}
\bibliography{ParticlePerformance}
\end{document}


\maketitle

\section{Memory Access Patterns for Contiguous Arrays}
\label{sec:supplemental:memaccess}

\begin{figure}[htbp]
  \centering
  \includegraphics[keepaspectratio, width=0.90\columnwidth]{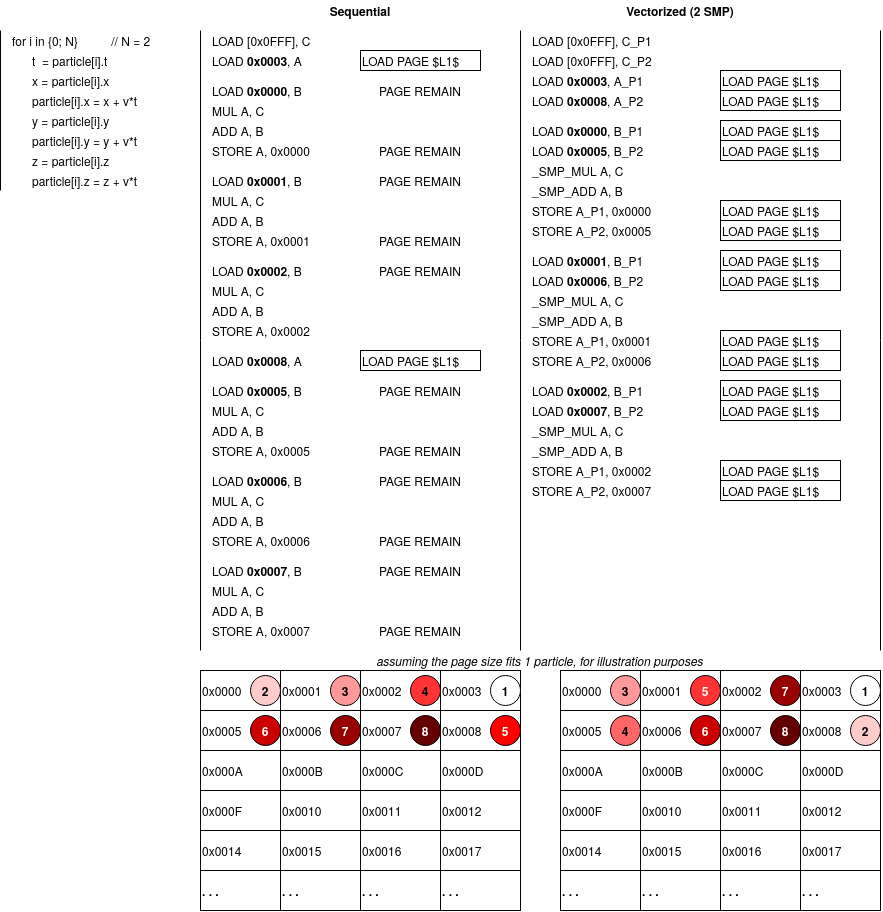}
  \caption{Memory access patterns related to a simple particle advection function on an $N x M$ array of structures, for sequential and vectorized evaluation (listed as assembly mnemonics).}
  \label{fig:appendix:memaccess:aos}
\end{figure}

\begin{figure}[htbp]
  \centering
  \includegraphics[keepaspectratio, width=0.90\columnwidth]{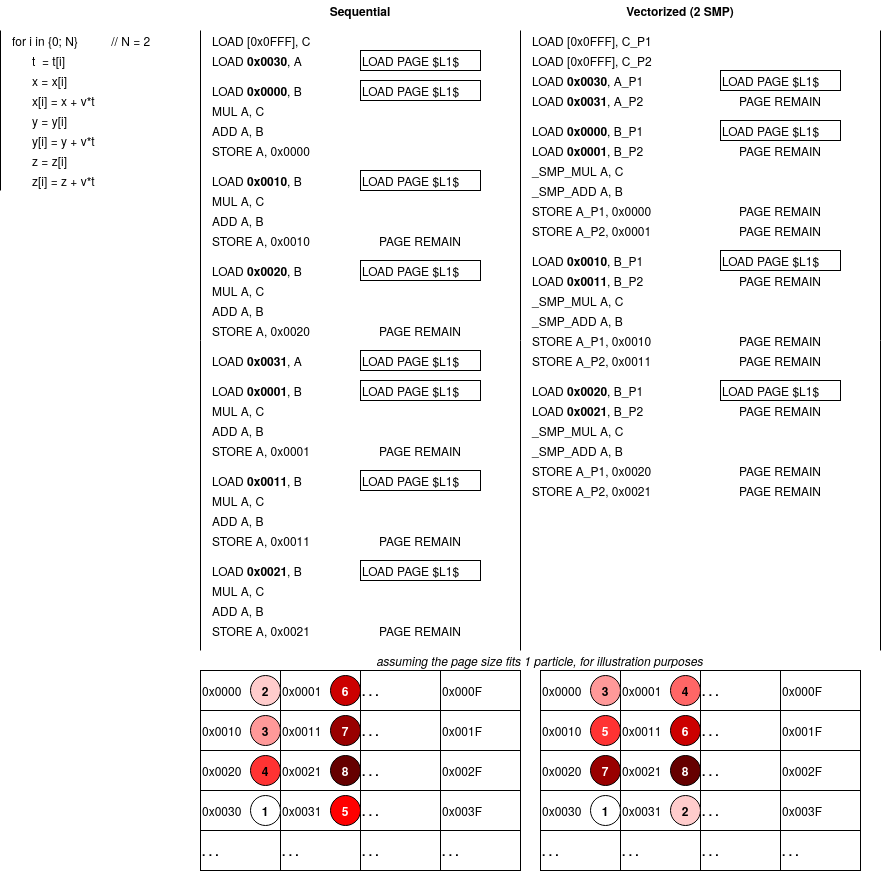}
  \caption{Memory access patterns related to a simple particle advection function on an $M x N$ structure of arrays, for sequential and vectorized evaluation (listed as assembly mnemonics).}
  \label{fig:appendix:memaccess:soa}
\end{figure}

$N x M$ arrays,  Arrays can be aligned in two patterns: in a usual and simple matter (1) the major axis represents the structure attributes aligned contiguously in memory, while the minor axis represents the multiple array items. Alternatively (2) the major axis can represent the array items contiguously aligned in memory, whereas the attributes follow the minor axis. Formally, with $N = |I|$, $I = \{x_{0}, x_{1}, ..., x_{n}\}$, and $M = |A|$ for $x(a)_{i} \forall a \in A$ (with $A$ representing or item structure \textit{attributes}), the arrays can structured as $N x M$ (1) or as (2) $M x N$. This has performance implications for the functional evaluation of the array, differing between a traditional sequential and a modern vectorized evaluation. This is shown in \cref{fig:appendix:memaccess:aos} for $N x M$ arrays (named \textit{Array of Structures (AoS)}, and in \cref{fig:appendix:memaccess:soa} for $M x N$ arrays (named \textit{Structure of Arrays (SoA)}.

Comparing the memory access patterns in terms of performance, we see that in traditional sequential processing the \textit{SoA} pattern is faster, as it only requires 1 cache update for each iteration whereas it requires 4 updates with an \textit{SoA} pattern. In modern computing architectures, each iteration actually executes between 2 (for multi-threaded CPUs) and 32 (for GPU warps) iterations simultaneously, thus a view on per-iteration (i.e. per-item) computations is invalid. This simultaneous computation is referred to as \textit{Vectorization} (on CPUs) or \textit{Simultaneous Multi-Processing (SMP)} (on general processors). Comparing the numbers for a 2-item evaluation, \textit{AoS} requires 2 cache updates and \textit{SoA} requires 8 cache updates when assuming a sequential processing. For vectorized processing though, evaluating 2 items requires 14 updates for \textit{AoS} and just 4 updates for \textit{SoA}. Furthermore, the number of cache changes is approximately $M \times \frac{N}{|PU|}$ (where $|PU|$ is the number of processing units), hence scaling favourably with the number of processing units and threads (in software). This naturally makes \textit{SoA} superior in performance for modern processors for $M << N$. Conversely, where $N \leq M$ one can just switch both array axes. Lastly, in practice, the layout change can be easily achieved by changing from FORTRAN-contiguous to C-contiguous order in Python arrays (or vice versa), which equally is achieved by matrix transposition. Obviously, establishing the correct item order on allocation and keeping the order static is vital, as a per-iteration matrix transposition consumes vastly more processing cycles than is saved by the \textit{SoA} evaluation.


 



\bibliographystyle{siamplain}
\bibliography{ParticlePerformance}